\documentclass[prl, twocolumn,showpacs,preprintnumbers,amsmath,amssymb ]{revtex4}

\usepackage{graphicx}
\usepackage{dcolumn}
\usepackage{bm}

\begin{document}

\input epsf.sty



\title{On the temperature dependence of the superconducting gap 
in high-$T_c$ cuprates}

\author{B. V. Fine}

\affiliation{
Max Plank Institute for the Physics of Complex Systems,
Noethnitzer Str. 38, D-01187, Dresden, Germany; and
\\
Physics Department, University of Tennessee, 1413 Circle Dr., Knoxville, TN 37996, USA}

\date{17 January, 2005}

\begin{abstract}
It is proposed that 
(i) the temperature dependence of the superconducting gap $\Delta(T)$
in high-$T_c$ cuprates can be  predicted just from the knowledge
of $\Delta(0)$ and the critical temperature $T_c$; and, in particular, 
(ii) $\Delta(0)/T_c > 4$ implies that $\Delta(T_c) \neq 0$, while 
$\Delta(0)/T_c \leq 4$ corresponds to $\Delta(T_c) = 0$.
A number of tunneling experiments appear to support
the above proposition, and, furthermore, show reasonable quantitative agreement
with a model based on the
two-dimensional stripe hypothesis.
\end{abstract}


\maketitle


In conventional superconductors, the phenomenology of the superconducting 
(SC) energy gap  
closely follows that of the model of Bardeen, Cooper and 
Schrieffer (BCS)\cite{BCS}. 
Namely: (i) Below the SC transition temperature ($T_c$),
there are no one particle excitations within the SC gap $\Delta(T)$, 
where $T$ is the temperature; (ii) At the gap boundaries $\pm \Delta(T)$, 
there are symmetric
SC peaks; (iii) The ratio $\Delta(0)/T_c$ 
is close to 1.76; (iv) $\Delta(T_c) = 0 $. 
In contrast, the tunneling studies of high-$T_c$ cuprates
show that: (i) The density of states within the SC gap is not zero;
(ii) The SC peaks are frequently asymmetric; (iii) The ratio $\Delta(0)/T_c$
is significantly larger than 1.76 and not universal; 
(iv) There are indications that, in some high-$T_c$ 
compounds, $\Delta(T_c) \neq 0$.
The above non-universality implies that any BCS-like
one-parameter description of the tunneling spectra cannot be
adequate for high-$T_c$ cuprates. The complexity of these materials 
further suggests that their proper microscopic description should contain
many parameters. 
In this work, however, I report a surprising finding
that the phenomenology
of the SC gap in high-$T_c$ cuprates may be describable by only two
parameters.

The present study was originally undertaken to test a simple model proposed in
Ref.~\cite{Fine-hitc-prb04}. 
The calculations of the tunneling characteristics based on that model
require only two input parameters $\Delta(0)$ and $T_c$. 
The reasonable success of this test, of course, 
lends credibility to the model, but also, irrespectively of the
model, it suggests that, even though 
the ratio $\Delta(0) / T_c$ may exhibit a non-universal doping dependence, 
and,
moreover, may be different for
different 
samples of the same material, the measured 
values of $\Delta(0)$
and $T_c$ are sufficient to predict the entire evolution of $\Delta(T)$.
Of particular interest is the following model-based rule:
\begin{equation}
\begin{array}{rcl}
\hbox{{\it $\ \Delta(0)/T_c > 4 \ \ \  \Leftrightarrow  \ \ \  \Delta(T_c) \neq 0 \ / $ asymmetric SC peaks;}  }
\\ 
\hbox{{\it $\ \ \Delta(0)/T_c \leq 4 \ \ \  \Leftrightarrow  \ \ \ \Delta(T_c) = 0 \ / $ \ symmetric SC peaks}. }
\end{array}
\label{rule}
\end{equation}
It was the surprisingly good agreement of the symmetry/asymmetry
part of this rule with many experiments\cite{Fine-hitc-prb04}, 
that prompted me to test the model predictions 
for the shape of the  $\Delta(T)$-curve.


\begin{figure} \setlength{\unitlength}{0.1cm}

\begin{picture}(100, 65) 
{ 
\put(13, 0){ \epsfxsize= 2.3in \epsfbox{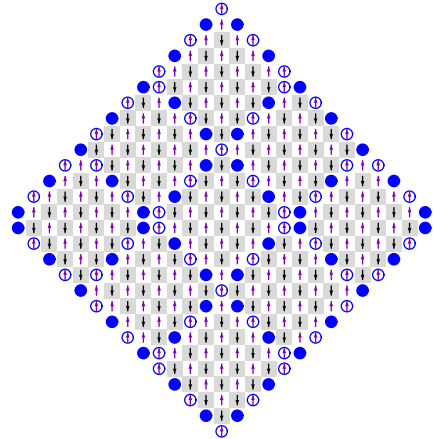} }
}
\end{picture} 
\caption{ 2D configuration of diagonal stripes (a cartoon).   
} 
\label{fig-stripes} 
\end{figure}


The model of Ref.~\cite{Fine-hitc-prb04} is based on the hypothesis 
that a two-dimensional (2D) arrangement of
stripes shown in Fig.~\ref{fig-stripes}  
exists in the CuO$_2$ planes of high-$T_c$ cuprates.
In such a superstructure, there exist hole states
localized either inside the stripes (b-states) or in the antiferromagnetic (AF) domains 
between the stripes (a-states). Due to the 
small size of the AF domains and stripe elements ($20-30$\AA), the
energy levels within each unit should be broadly spaced 
(estimated\cite{Fine-hitc-prb04} as 40~meV). 
The model proceeds by assuming the existence of only one low-energy a-state
per AF domain  and
two low-energy b-states per stripe element. 
As argued in Ref.~\cite{Fine-hitc-prb04}, a promising choice 
for the model Hamiltonian is
\begin{eqnarray}
{\cal H} &=&\varepsilon_a \sum_i a_i^+ a_i \ + \ 
\varepsilon_b \sum_{i, j(i), \sigma}^{\eta_i= 1} b_{ij,\sigma}^+ b_{ij,\sigma} 
\nonumber \\ 
&& + \ g \sum_{i, j(i)}^{\eta_i= 1} ( b_{ij,+}^+ b_{ij,-}^+ a_i a_j + \hbox{h. c.}),
\label{H}
\end{eqnarray}
where index $i$ or $j$ labels AF domains; 
$j(i)$ implies, that the $j$th AF domain is the nearest neighbor
of the $i$th domain; $a_i$ is the annihilation 
operator of a hole inside the $i$th AF domain;
$b_{ij,\sigma}$ is the annihilation operator of a hole inside 
the stripe element separating
the $i$th and the $j$th AF domains; $\sigma$ is the spin index, which
can have two values ``$+$'' or ``$-$''; 
$\varepsilon_a$ and $\varepsilon_b$
are position-independent 
on-site energies for a- and b-states, respectively, counted from the
chemical potential; and, finally, $g$ is the coupling constant.
The spins of a-states are tracked by index $\eta_i$, which 
alternates between $1$ or $-1$ thus tracing the sign of the AF
order parameter in the AF domains. 
The sum superscript ``$\eta_i =1$''  
indicates that the summation extends only
over the supercells having $\eta_i =1$.

The total energy of such a
system as a function of the chemical potential $\mu$ has at 
least two minima\cite{Fine-hitc-prb04}:
$\mu =\varepsilon_a$ and $\mu = \varepsilon_b$,
which implies two most promising 
regimes --- 
Case~I: $\varepsilon_b =0 $; and
Case~II: $\varepsilon_a = 0 $. 
The mean-field solution  then 
leads to the density of states 
described by the following system of equations\cite{Fine-hitc-prb04}:
in Case~I,
\begin{equation}
\varepsilon_A({\mathbf{k}}) = \sqrt{\varepsilon_a^2 + 
{1 \over 4} g^2 (2 n_B -1)^2 |V({\mathbf{k}})|^2
},
\label{epsAkI}
\end{equation}
\begin{equation}
\varepsilon_B = - {g^2 (2 n_B -1) \over 64 \pi^2}
\int_{-{\pi}}^{{\pi}} dk_x 
\int_{-{\pi}}^{{\pi}} dk_y 
{(1 - 2 n_A({\mathbf{k}}) |V({\mathbf{k}})|^2
\over \varepsilon_A({\mathbf{k}})},
\label{epsBI}
\end{equation}
and, in Case II,
\begin{eqnarray}
\varepsilon_A({\mathbf{k}}) &=&
- {g^2 \ |V({\mathbf{k}})| \ C_a \ (1 - 2 n_B) \over 8 \varepsilon_B},
\label{epsAkII}
\\
\varepsilon_B &=& \sqrt{\varepsilon_b^2 + 
{1 \over 16} \ g^2 \ C_a^2
},
\label{epsBII}
\end{eqnarray}
where  subscripts $A$ and $B$ indicate the Bogoliubov quasiparticles
associated with a- and b-states, respectively;
$\varepsilon_{A/B}$ and $n_{A/B}$ are the quasiparticle energies
and the occupation numbers, respectively.  A-quasiparticles are 
characterized by well-defined wave vectors $\mathbf{k}$.
All B-quasiparticles have the same  $\varepsilon_B$.
The expressions for $n_A({\mathbf{k}})$ and $n_B$, and
for the auxiliary quantities
$C_a$ and $V({\mathbf{k}})$ are:
\begin{equation}
n_A({\mathbf{k}}) = {1 \over \hbox{exp}\left({\varepsilon_A({\mathbf{k}}) \over T}\right) +1 },
\label{nAk}
\end{equation}
\begin{equation}
n_B = {1 \over \hbox{exp}\left({\varepsilon_B \over T}\right) +1 },
\label{nB}
\end{equation}
\begin{equation}
V({\mathbf{k}}) = 2 \left[
\hbox{cos} \left( {k_x +k_y \over 2}  \right)  -
i \  \hbox{cos} \left( {k_x -k_y \over 2}  \right) 
\right],
\label{Vcc}
\end{equation}
\begin{equation}
C_a = {1 \over 8 \pi^2} 
\int_{-{\pi}}^{{\pi}} dk_x 
\int_{-{\pi}}^{{\pi}} dk_y 
(2 n_A({\mathbf{k}}) -1) |V({\mathbf{k}})|. 
\label{Ca}
\end{equation}
The tunneling spectra of A-states have Van Hove singularities, 
which, in Case I, are located
at 
\begin{equation}
\varepsilon_{A0} = \pm \ \sqrt{
\varepsilon_a^2 +   g^2 (2 n_B -1)^2 
}.
\label{epsA0I}
\end{equation}
and, in Case II, at
\begin{equation}
\varepsilon_{A0} = 
\pm \ {g^2 C_a (1 - 2 n_B) \over 4 \varepsilon_B}
\label{epsA0II}
\end{equation}
Three representative zero-temperature tunneling spectra of both
A- and B-states can be found in Ref.~\cite{Fine-hitc-prb04}.

In both Cases I and II, the mean-field $T_c$ 
can be obtained~\cite{Fine-hitc-prb04} numerically from 
the following equation:
\begin{equation}
T_c = { g^2 \left[\hbox{exp}\left({|\varepsilon_a - \varepsilon_b| \over T_c}\right) - 1\right] \over 
8 |\varepsilon_a - \varepsilon_b| 
\left[\hbox{exp}\left({|\varepsilon_a - \varepsilon_b| \over T_c}\right) + 1\right]}.
\label{Tc}
\end{equation}

Given the above picture, the basic features of the tunneling phenomenology
can be interpreted as follows:
(1)~The SC peaks at
$\pm \Delta$, are identified with
the Van Hove singularities at $\pm \varepsilon_{A0}$ in the density of A-states.
(2)~The density of B-states is assumed to be more difficult to detect due to,
perhaps, smaller tunneling elements. 

\begin{figure}
\setlength{\unitlength}{0.1cm}
\begin{picture}(100, 68)
{
\put(-3, 0){
\epsfxsize=3.5in
\epsfbox{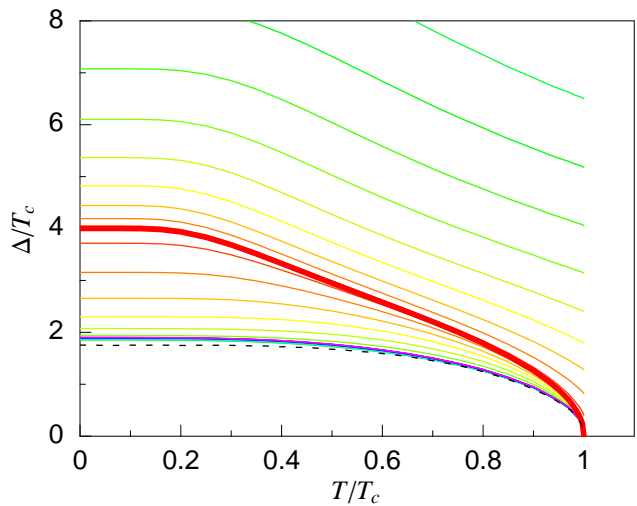} }
}
\end{picture} 
\caption{(Color online) Family of theoretical curves for the 
temperature dependence
of the superconducting gap. Thick line corresponds to the critical ratio
$\Delta(0)/T_c = 4$. Solid lines above the thick line describe
Case I, and below the thick line  Case II.
The dashed line represents the BCS result.
} 
\label{fig-family} 
\end{figure}

The knowledge of $\Delta(0)$ and $T_c$ is thus sufficient to obtain the 
model parameters $|\varepsilon_a|$ and $g$ in Case~I 
[Eqs.(\ref{epsA0I},\ref{Tc}) with $n_B = 1$], and 
$|\varepsilon_b|$
and $g$ in Case II 
[Eqs.(\ref{epsBII},\ref{Vcc},\ref{Ca},\ref{epsA0II},\ref{Tc}) with 
$n_A({\mathbf{k}}) = 1; n_B = 0$]. 
After that, $\Delta(T)$ can be calculated numerically 
using Eqs.(\ref{epsAkI}-\ref{epsA0II}) without additional
adjustable parameters. 
Importantly, Cases I and II do not overlap in terms of
the ratio $\Delta(0)/T_c$: in Case I, $\Delta(0)/T_c > 4$, while in Case II,
$\Delta(0)/T_c < 4$.
The entire family of the $\Delta(T)$-curves is presented in Fig.~\ref{fig-family}.


In both Cases I and II, $|\varepsilon_a - \varepsilon_b|$ is a measure
of the pseudogap (see Sec.V of Ref.~\cite{Fine-hitc-prb04}). This pseudogap is a real feature 
in the normal  density of states (as, e.g., in the proposal of Ref.\cite{Loram-etal-94}), 
i.e., it is not associated with the SC fluctuations 
above $T_c$.  This feature is a generic 
consequence of the nanoscale phase separation ---
two different kinds of real space regions (stripes and AF domains) 
imply two different energies associated with them.
The model pseudogap equals zero only in the critical case,
$\varepsilon_a = \varepsilon_b$, which, as argued in 
Ref.\cite{Fine-hitc-prb04}, may correspond to 
the doping concentration 0.19.
The model has two different SC energy gaps: 
$\varepsilon_{A0}$ for A-states and $\varepsilon_B$ for B-states.
In Case I, $\varepsilon_B$ turns to zero at $T=T_c$, while
$\varepsilon_{A0}$ becomes equal to the pseudogap 
$|\varepsilon_a - \varepsilon_b|$.
In Case II, the situation at $T =T_c$ is opposite:
$\varepsilon_{A0} =0$, while 
$\varepsilon_B =|\varepsilon_a - \varepsilon_b|$.
Thus the difference between the two model regimes
is not qualitative but only quantitative. 
However, since, by assumption, the observable SC peaks
represent only A-states, the phenomenology of the SC gap appears
to be qualitatively different in the two regimes.

\begin{figure*}
\setlength{\unitlength}{0.1cm}
\begin{picture}(200, 55)
{
\put(-25, 0){
\epsfxsize=8.75in
\epsfbox{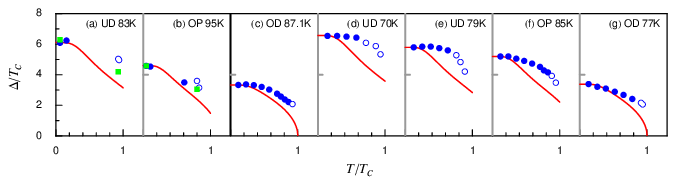} }
\put(86,50){\textsf{\large Bi-2212}}
\put(15,45){\textsf{(a,b) BJ - Miyakawa et al.,  Ref.\cite{Miyakawa-etal-98} }}
\put(70,45){\textsf{(c) ILT - Suzuki et al., Ref.\cite{SWM}  \ \ \ \ \
                                 (d-g) - Suzuki \& Watanabe, Ref.\cite{SW} }}
}
\end{picture} 
\begin{picture}(200, 50)
{
\put(-25, 0){
\epsfxsize=8.75in
\epsfbox{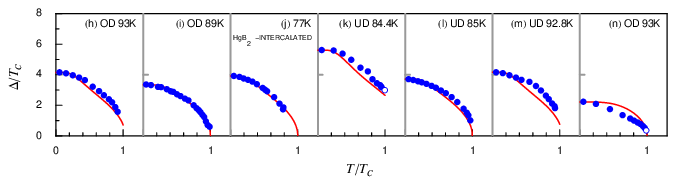} }
\put(15,45){\textsf{(h,i) ILT - Krasnov et al.,  
Ref.\cite{Krasnov-etal-00} \ \ \ \ \ 
(j) ILT - Krasnov, Ref.\cite{Krasnov-02A} \ \ \ \ \
(k-n) ILT - Krasnov, Ref.\cite{Krasnov-02} }}
}
\end{picture} 
\begin{picture}(200, 57)
{
\put(-25, 0){
\epsfxsize=8.75in
\epsfbox{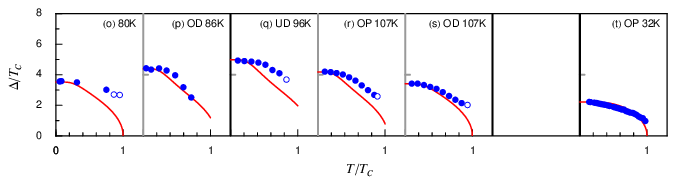} }
\put(30,55){\textsf{\large Bi-2212}}
\put(87,50){\textsf{\large Bi-2223}}
\put(144,50){\textsf{\large Bi-2201-La(0.4)}}
\put(15,50){\textsf{(o) BJ - Vedeneev et al.,  Ref.\cite{Vedeneev-etal-94} }}
\put(15,45){\textsf{(p) BJ - Akimenko et al.,  Ref.\cite{Akimenko-etal-99} }}
\put(65,45){\textsf{(q-s) ILT - Yamada et al., Ref.\cite{Yamada-etal-03} }}
\put(127,45){\textsf{(t) ILT - Yurgens et al., Ref.\cite{Yurgens-etal-03} }}
}
\end{picture} 
\caption{(Color online) Temperature evolution of the SC gap.
Solid lines - theory; circles - experiments;
BJ - break junction; ILT - interlayer tunneling. 
Horizontal marks in each frame
indicate $\Delta/T_c = 4$. 
Whenever the original tunneling spectra were reported 
(a-h, k, n-t), filled circles indicate reasonably well-pronounced SC peaks,
while open circles indicate very broad
{\it and} small SC peaks. Filled circles in frames (i,j,l,m)
imply no additional information about the SC peaks. Squares
in frames (a,b) represent the ``bare'' values of $\Delta$
estimated in Ref.\cite{Miyakawa-etal-98}.
} 
\label{fig-exper} 
\end{figure*}

In Fig.~\ref{fig-exper}, I compare the theoretical predictions
with the results of the break junction 
(BJ)\cite{Vedeneev-etal-94,Miyakawa-etal-98,Akimenko-etal-99} 
and  the interlayer tunneling 
(ILT)\cite{SWM,SW,Krasnov-etal-00,Krasnov-02,Yurgens-etal-03,Yamada-etal-03}
experiments. 
The ILT technique gives better
energy resolution and comes closer to measuring the true intrinsic
density of states in the bulk of cuprates, but, at the same time, 
it generates significant overheating of the sample.
In order to avoid this problem, the authors of 
Refs.\cite{SWM,SW,Yamada-etal-03}  
have used a short-pulse
method, which, however, comes at a cost of somewhat lower energy resolution
[Figs.~\ref{fig-exper}(c-g) and (q-s)].
Alternatively, the authors of Refs.\cite{Krasnov-etal-00,Krasnov-02,Krasnov-02A,Yurgens-etal-03} 
[Figs.~\ref{fig-exper}(h-n) and (t)] have perfected 
the ``conventional'' ILT technique 
by manufacturing sufficiently small {\it mesas} and thus 
reducing the overheating.
In the latter case, the experimental data points reflect 
the ambient temperature of experiment. The 
question, however, remains concerning the true value of the temperature
of the {\it mesas}\cite{Zavaritsky,Zavaritsky-04}.
Theoretical and experimental estimates made 
in Refs.\cite{Krasnov-02A,Yurgens-etal-03A,Krasnov-etal-04}, 
have indicated possible significant overheating,
but there is still a chance
that, at least in the underdoped Bi-2212, the temperature 
was not too much distorted  at the voltage biases probing the
SC peaks.


Most of the theoretical curves shown in Fig.~\ref{fig-exper} 
agree with the experimental data
within the limits of [sometimes, large]
uncertainty associated with the broadening of the SC peaks 
(see the original references).
The data points in Fig.~\ref{fig-exper} are obtained
from the peak-to-peak separation in the experimental tunneling spectra.  
Usually, the peak broadening shifts  
the peak maximum towards higher energies. 
If this effect is undone by extracting the ``bare''
values of $\Delta$, then the experimental points in 
Figs.~\ref{fig-exper}(a-g,o,q-s)
would appear significantly closer to the theoretical curves. This is 
a consequence of the fact that the 
SC peaks are broader at higher temperatures.
The magnitude
of the above correction can be judged on the basis of 
the estimates 
of bare $\Delta$
made in Ref.\cite{Miyakawa-etal-98} and 
presented in Fig.~\ref{fig-exper}(a,b). [Note: the widths of the SC
peaks corresponding to Figs.~\ref{fig-exper}(a,b) are comparable or smaller
than those corresponding to Figs.~\ref{fig-exper}(d-g,o-s).]

The agreement between 
the theory and the experiment is typically better, 
whenever the SC peaks are sharper.
This is, particularly,
true for the ``pulsed'' ILT experiments (Fig.~\ref{fig-exper}(c-g) and (q-s)):
up to $T/T_c \sim 0.7$ the data points from the 
overdoped samples [Figs.~\ref{fig-exper}(c,g,s)] correspond to 
sharper SC peaks and   
agree better with the theory.
Figure~\ref{fig-exper}(n) constitutes an exception from the above trend, but,
in this case, a stronger degree of overheating is suspected.
The best agreement between the theory and the experiment 
can be observed in Figs.~\ref{fig-exper}(h-m),
in which case the data points were generated by the ``conventional''
ILT technique (expected to give the best energy resolution). 
If these data also stand the test of time with respect
to the true temperature of the samples, then  such an agreement
would amount to a striking success of the present model.

The experimental data in Figs.~\ref{fig-exper} are also fairly
consistent with the first line of the rule~(\ref{rule}). 
As far as the second line of that rule is concerned, then 
Figs.~\ref{fig-exper}(i) and (n) can 
be cited as supporting the rule. 
The data points in Figs.~\ref{fig-exper}(c,g,o,s) while corresponding to
$\Delta(0)/T_c < 4$ show trend towards $\Delta(T_c) \neq 0$. However
this trend is due to the highest temperature points, which originate
from poorly characterized SC peaks. The conclusive verification
of rule (\ref{rule}) should thus await further experiments. 
It should also be noted in this context, that the 
scanning tunneling spectroscopy studies of Bi-2212\cite{Renner-etal-98} and
Bi-2201\cite{Kugler-etal-01}, which indicated that 
$\Delta(T_c) \neq 0$, were all performed on the samples exhibiting
$\Delta(0)/T_c > 4$.

In a future study, I plan to test the model predictions for the specific
heat experiments, which should be equally sensitive to the densities of both $A$- and $B$-states.

In conclusion, I express cautious optimism that 
(i)~there exists a two-parametric
description of the  $\Delta(T)$-curves in high-$T_c$ cuprates;
(ii) there exists a critical ratio $\Delta(0)/T_c $, which
signifies a transition between two different kinds of SC states, 
and (iii) the above critical ratio is approximately equal to 4.
The quantitative agreement between the experimental data 
reproduced in Fig.~\ref{fig-exper}(h-m) and the
model calculations also appears quite
promising. 

The author is grateful to V. M. Krasnov, A. Yurgens, N. Miyakawa,
M. Suzuki, and V.~N.~Zavaritsky for valuable discussions.

\bibliography{hitc3}

\end{document}